\documentclass[aps,prl,reprint,superscriptaddress,floatfix,10pt]{revtex4-1} 
\usepackage{graphicx}
\usepackage{amssymb,amsmath}
\usepackage{color}
\usepackage{hyperref}
\usepackage{etoolbox}
\newcommand{\supplement}{Supplemental Material}

\allowdisplaybreaks[2]

\begin{document}

\title{Tunable sub-luminal propagation of narrowband x-ray pulses}

\author{Kilian P. Heeg}
\affiliation{Max-Planck-Institut f\"ur Kernphysik, Saupfercheckweg 1, 69117 Heidelberg, Deutschland}

\author{Johann Haber}
\affiliation{Deutsches Elektronen-Synchrotron DESY, Notkestra{\ss}e 85, 22607 Hamburg, Germany}

\author{Daniel Schumacher}
\affiliation{Deutsches Elektronen-Synchrotron DESY, Notkestra{\ss}e 85, 22607 Hamburg, Germany}

\author{Lars Bocklage}
\affiliation{Deutsches Elektronen-Synchrotron DESY, Notkestra{\ss}e 85, 22607 Hamburg, Germany}

\author{Hans-Christian Wille}
\affiliation{Deutsches Elektronen-Synchrotron DESY, Notkestra{\ss}e 85, 22607 Hamburg, Germany}

\author{Kai S. Schulze}
\affiliation{Institut f\"ur Optik und Quantenelektronik, Friedrich-Schiller-Universit\"at Jena, Max-Wien-Platz 1, 07743 Jena, Germany}
\affiliation{Helmholtz-Institut Jena, Fr\"obelstieg 3, 07743 Jena, Germany}

\author{Robert Loetzsch}
\affiliation{Institut f\"ur Optik und Quantenelektronik, Friedrich-Schiller-Universit\"at Jena, Max-Wien-Platz 1, 07743 Jena, Germany}
\affiliation{Helmholtz-Institut Jena, Fr\"obelstieg 3, 07743 Jena, Germany}

\author{Ingo Uschmann}
\affiliation{Institut f\"ur Optik und Quantenelektronik, Friedrich-Schiller-Universit\"at Jena, Max-Wien-Platz 1, 07743 Jena, Germany}
\affiliation{Helmholtz-Institut Jena, Fr\"obelstieg 3, 07743 Jena, Germany}

\author{Gerhard G. Paulus}
\affiliation{Institut f\"ur Optik und Quantenelektronik, Friedrich-Schiller-Universit\"at Jena, Max-Wien-Platz 1, 07743 Jena, Germany}
\affiliation{Helmholtz-Institut Jena, Fr\"obelstieg 3, 07743 Jena, Germany}

\author{Rudolf R\"uffer}
\affiliation{ESRF-The European Synchrotron, CS40220, 38043 Grenoble Cedex 9, France}

\author{Ralf R\"ohlsberger}
\affiliation{Deutsches Elektronen-Synchrotron DESY, Notkestra{\ss}e 85, 22607 Hamburg, Germany}

\author{J\"org Evers}
\affiliation{Max-Planck-Institut f\"ur Kernphysik, Saupfercheckweg 1, 69117 Heidelberg, Deutschland}

\date{\today}

\begin{abstract}
Group velocity control is demonstrated for x-ray photons of 14.4~keV energy via a direct measurement of the temporal delay imposed on spectrally narrow x-ray pulses. Sub-luminal light propagation is achieved by inducing a steep positive linear dispersion in the optical response of $^{57}$Fe M\"ossbauer nuclei embedded in a thin film planar x-ray cavity. The direct detection of the temporal pulse delay is enabled by generating frequency-tunable spectrally narrow x-ray pulses from broadband pulsed synchrotron radiation. Our theoretical model is in good agreement with the experimental data.
\end{abstract}

\maketitle
Strong nonlinear interaction of light with matter is a key requirement for fundamental and applied quantum optical technologies alike. Since conventional materials typically exhibit weak nonlinearities, the ultimate quest for strong nonlinear interactions of individual quanta has led to the development of a number of methods to significantly enhance nonlinear light-matter interactions. Among the most prominent ones are coherently prepared media based on electromagnetically induced transparency, sub-luminal light and related effects~\cite{Fleischhauer2005,focus}, as well as cavity-enhanced light matter interactions~\cite{cqed}. 

Recently, nuclear quantum optics featuring the interaction of x-ray light with M\"ossbauer nuclei in the few keV transition energy range has gained considerable momentum, both theoretically~\cite{PhysRevLett.82.3593,PhysRevLett.82.3593,Palffy2009,Buervenich2006,tenBrinke2013,Liao2012,Heeg2013b} and experimentally~\cite{Shvydko1996,Coussement2002,Shakhmuratov2009,Roehlsberger2010,Shakhmuratov2011,Roehlsberger2012,
Heeg2013,Vagizov2014,Adams2003,Adams2013}.
Interestingly, these experiments operate with less than one resonant x-ray photon per pulse on average due to restrictions in the available x-ray light sources. This raises the question, whether coherent or cavity-based enhancement techniques could be utilized to realize nonlinear light-matter interactions in nuclear quantum optics despite the low number of resonant photons. 

Here, we report a first step towards this goal, and demonstrate group velocity control of spectrally narrow x-ray pulses (SNXP). Sub-luminal light propagation is achieved by inducing a steep positive linear material dispersion, and verified by direct measurements of the temporal delay imposed on the SNXP. For this, we suitably manipulate the optical response of the $\omega_0=14.4$ keV M\"ossbauer resonance (single nucleus linewidth $\gamma=4.7$ neV) of a large ensemble of $^{57}$Fe nuclei embedded in a thin film planar x-ray cavity. Our approach thereby combines coherent control, as well as cooperative and cavity enhancements of light-matter interaction in a single setup. To enable the direct detection of the temporal pulse delay, we further propose and implement a flexible scheme to generate frequency-tunable SNXP from broadband synchrotron radiation for applications in x-ray quantum optics. Our theoretical model is in good agreement with the experimental data.

Sub-luminal light was first demonstrated in the visible frequency range~\cite{Hau1999,Kash1999,Budker1999}, and by now has been implemented in a number of platforms~\cite{focus}, particularly also in cavity settings~\cite{ISI:000278551800039,ISI:000295397400018}. Manipulation of light propagation has also been reported in the x-ray regime. In Ref.~\cite{Shakhmuratov2009}, a delayed peak in the transmitted x-ray light intensity has been observed. In this case, however, the pulse delay is induced by the propagation of the light through a doublet absorber structure rather than electromagnetically induced transparency or related effects, and can be interpreted as arising from transitions between super- and subradiant states. Also coherent storage of light via rapid control of the applied quantization field has been achieved~\cite{Shvydko1996}. Other experiments with nuclei observed electromagnetically induced transparency~\cite{Roehlsberger2012}, related spontaneously generated coherences with equivalent susceptibilites~\cite{Heeg2013}, or other transparency mechanisms~\cite{Coussement2002,Odeurs2010}. However, these experiments did not study the delay or the actual pulse propagation. 
In contrast, in our experiment, we induce slow light via a steep linear dispersion, and verify the x-ray group velocity control via a direct observation of the temporal pulse delay.

\begin{figure}[t]
 \centering
 \includegraphics[scale=0.8]{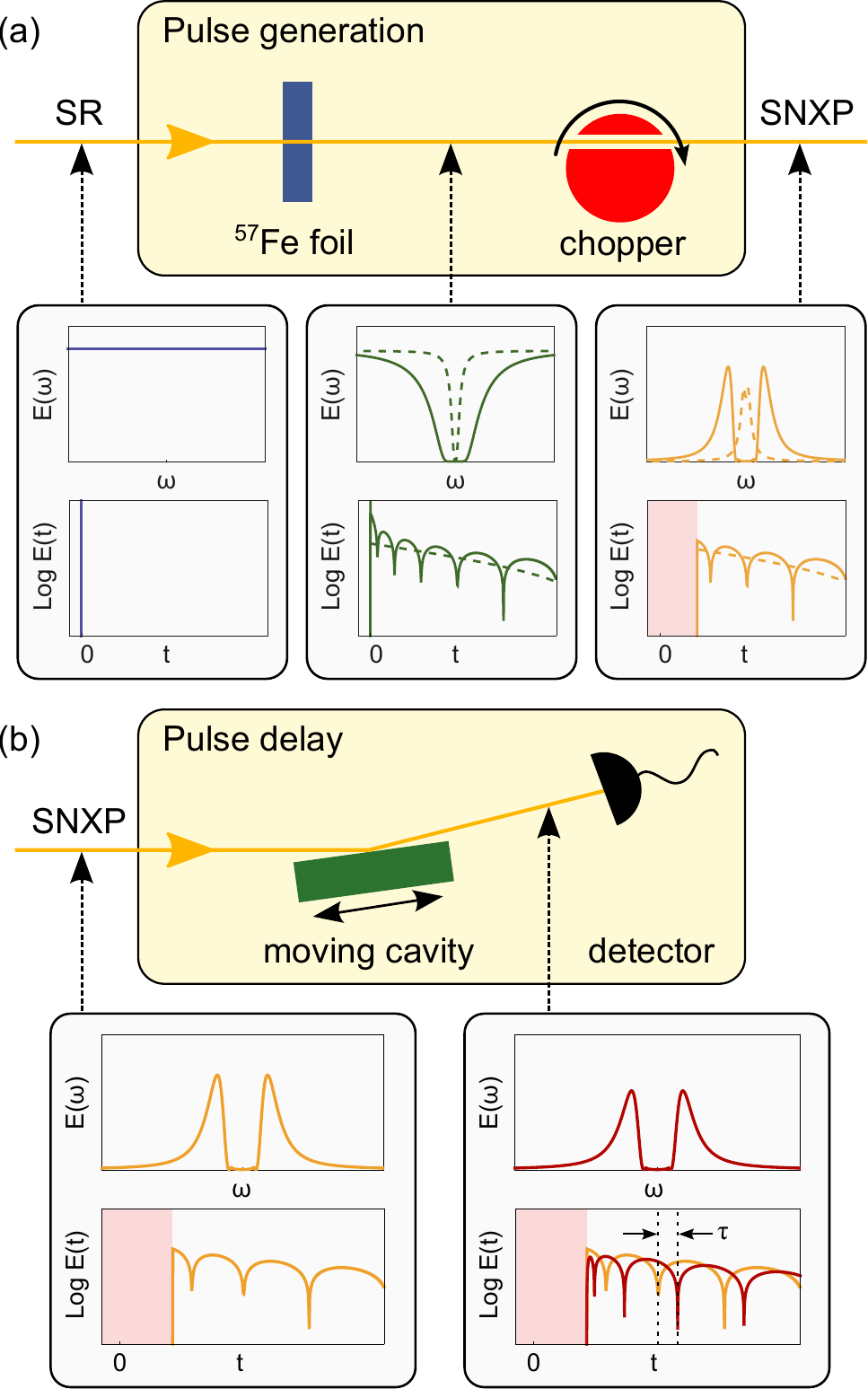}
 \caption{\label{fig:setup}(Color online) (a) Possible setup for the generation of SNXP from broadband SR. A $^{57}$Fe foil imprints an absorption band on the initially flat spectrum. This absorption is converted into a SNXP by suppressing the unscattered prompt response, e.g., via a mechanical chopper. The lower panels depict the frequency- and time-resolved amplitude of the field throughout the generation for a thick (solid lines) and thin (dashed lines) $^{57}$Fe foil. The red shaded areas indicate the time gating by the chopper. (b) X-ray group velocity control. A SNXP is reflected off of a thin-film cavity containing near-resonant nuclei. The nuclear dispersion imprints a phase shift onto the pulse, which results in a delay of the pulse without distortion. The delay is most clearly visible in comparing the beating minima in the time domain, which arise from the double-pulse structure in the frequency spectrum of the input pulse. The two curves in the bottom right panel show the pulse with and without the induced delay $\tau$.}
\end{figure}

\begin{figure}[t]
 \centering
 \includegraphics[scale=0.8]{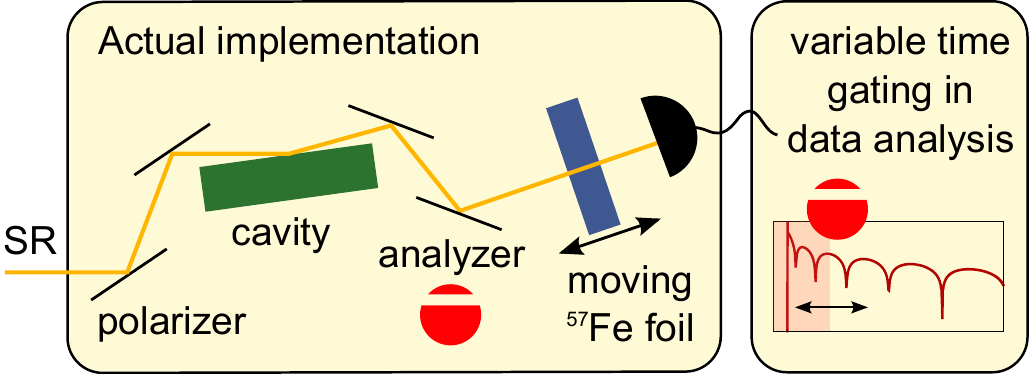}
 \caption{\label{fig:implementation}(Color online) Actual implementation of the experiment. A polarimeter blocks the background photons such that neither mechanical chopper nor a high-resolution monochromator for the SR are required. The variable a posteriori time gating facilitates the analysis of the delayed SNXP. The red circles indicate which elements replace the chopper in Fig.~\ref{fig:setup}.}
\end{figure}

{\it SNXP generation.} The desired group velocity control and subsequent applications require a SNXP as input field, such that the linear part of the nuclear dispersion covers the SNXP spectrum. In the x-ray regime, narrow-band radiation is provided by M\"ossbauer radioactive sources, but they are not pulsed, except for scenarios where special modulation schemes are applied~\cite{Vagizov2014}. Pulsed x-rays are preferably provided by synchrotron radiation (SR) sources, where nuclear resonant spectroscopy is an established method~\cite{Roehlsberger2005}. The technique relies on broadband excitation of nuclear levels and subsequent detection of the delayed nuclear decay signal. However, narrowband filtering of a single line from SR with sufficient rejection ratio is challenging since the beam has a bandwidth orders of magnitude larger than the nuclear resonance. One approach in this direction has been recently successfully demonstrated~\cite{Smirnov1997,Mitsui2007,Potapkin2012}. In this case, a narrowband, pure nuclear reflection from a $^{57}$FeBO$_3$ crystal is employed to suppress the enormous fraction of nonresonant photons in the incident beam. Another approach relies on a high-speed mechanical chopper~\cite{Toellner2011}. In this method, a $^{57}$Fe foil adds a tail of delayed narrow-band light scattered by the nuclei to the x-ray pulse, as shown in Fig.~\ref{fig:setup}(a). The chopper is operated such that it blocks the temporally short broadband incident pulse, but lets the delayed signal pass. As a result, the spectral dip induced by the iron foil is converted into a SNXP, which can then be used in the actual experiment. The characteristics of the generated SNXP spectrum are determined by the thickness of the iron foil. A thin or less enriched foil results in a single peak, whereas for enriched thicker foils the double-hump distribution well-known from nuclear resonance scattering is created~\cite{buerck1999}. These two cases are illustrated as dashed and solid lines, respectively, in the lower part of Fig.~\ref{fig:setup}.
In our experiment, we generate the SNXP using a equivalent method based on polarization filtering~\cite{Toellner1995,Marx2011,Heeg2013} for the suppression of the background photons. This implementation shown in Fig.~\ref{fig:implementation} does not require mechanical choppers, and is discussed in more detail below.

{\it Group velocity control.} The basic setup for the group velocity control is illustrated in Fig.~\ref{fig:setup}(b). The nuclei inducing the steep linear dispersion are embedded in a nm-sized x-ray cavity. A SNXP is reflected off this cavity in grazing incidence, and the reflected light is subsequently detected. 
The input SNXP is characterized by $E(\omega)$ in the frequency domain and by the Fourier transform $E_1(t) \propto \int E(\omega) e^{-i\omega t} d\omega$ in the time domain. The cavity modifies the SNXP spectrum via its complex reflection coefficient $R_\textrm{Cavity}(\omega)$, which, for instance, can be calculated with the quantum optical model introduced in Ref.~\cite{Heeg2013b} (for details see the \supplement). Since the pulse $E(\omega)$ is spectrally narrow around its center frequency $\omega_0$ compared to the superradiantly broadened cavity reflectance, we can approximate
$R_\textrm{Cavity}(\omega) \approx R_\textrm{Cavity}(\omega_0) \exp{[ i (\omega-\omega_0) \tau]}$, 
where
\begin{align}
\tau = \tfrac{\partial}{\partial \omega} \arg [ R_\textrm{Cavity}(\omega_0)]\,,
\end{align}
such that the temporal response of the setup is given by
\begin{align}
 E_2(t) &\propto \int E(\omega) R_\textrm{Cavity}(\omega) e^{-i\omega t} d\omega \nonumber \\
 &\approx R_\textrm{Cavity}(\omega_0) e^{ -i \omega_0 \tau}\: \int E(\omega) e^{-i\omega (t-\tau)} d\omega \nonumber \\
 &\propto  E_1(t - \tau) \label{eq:E_2}\;.
\end{align}
We thus find that the SNXP is delayed by the time $\tau$ due to the cavity dispersion without distortion of the pulse shape, as it is well known from cavities and waveguides in the visible regime~\cite{Longhi2002,Heebner2002}. The group delay $\tau$ can be tuned via a Doppler shift induced by moving the cavity, such that the cavity spectrum $R_\textrm{Cavity}(\omega)$ is detuned with respect to the pulse spectrum $E(\omega)$. In the cavity setting, the delay is determined by the slope of $\arg(R_\textrm{Cavity})$, whereas in atomic gases the real part of the susceptibility $\chi$ takes this role. The relation between the complex reflection coefficient $R_\textrm{Cavity}$ and the susceptibility is discussed in more detail in the \supplement.

{\it Experimental implementation.}
As already mentioned, the generation of the SNXP in our scheme avoids using a mechanical chopper. Instead, the initial broadband SR pulse is directed into a high purity x-ray polarimeter \cite{Marx2011}, see Fig.~\ref{fig:implementation}. The cavity containing the nuclei is placed between its polarizer and analyzer. After the polarimeter, the x-rays pass the $^{57}$Fe foil, and are subsequently detected by an avalanche photo diode. 
Compared to the setup in Fig.~\ref{fig:setup}, in our scheme, the order of the cavity and the $^{57}$Fe foil are reversed. This is possible, since all responses are linear. Second, instead of moving the cavity to tune the group delay, the $^{57}$Fe foil is moved, which is easier to realize and equivalent via a change of reference frame since both the source and the detection are spectrally broad. Most importantly, the mechanical chopper essential to the setup in Fig.~\ref{fig:setup} is not required in our scheme, since the polarimeter is operated in crossed setting. Thus, only those photons arrive at the detector, whose polarization has been rotated by the interaction with the nuclei. Thereby, the non-resonant background is removed, such that no high-resolution monochromator for the incident SR pulse is required, and the remaining signal can be detected without time gating. Apart from the simplification of the experimental setup, this also opens the possibility to a posteriori choose arbitrary time gatings in the data analysis. This is of interest, since after the $^{57}$Fe foil, the resulting detected signal becomes $R(t) = R_\delta(t) + R_\textrm{SNXP}(t)$, where $R_\delta(t)$ corresponds to photons which passed the $^{57}$Fe foil without interacting, and $ R_\textrm{SNXP}(t)$ to photons which did interact (see \supplement for details). The desired delayed part $R_\textrm{SNXP}(t)$ can be separated from $R_\delta(t)$ by time gating. Thus, our approach allows to optimize this time gating throughout the data analysis.

\begin{figure}[t]
 \centering
 \includegraphics[width=\columnwidth]{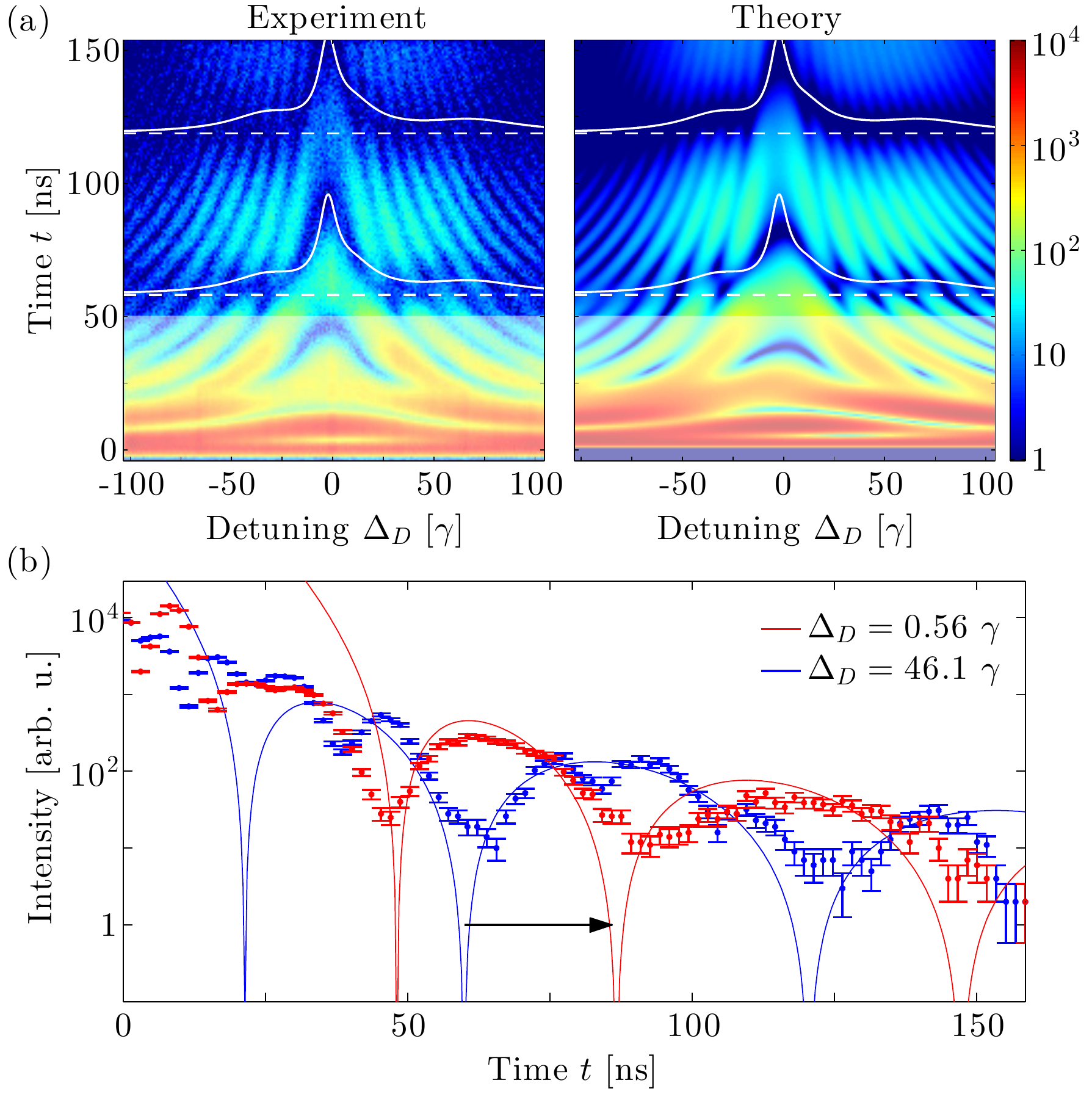}
 \caption{\label{fig:counts}(Color online) (a) Photon counts as function of time and Doppler detuning of the $^{57}$Fe foil. White dashed lines indicate theoretical predictions for beating minima positions without pulse delay. Solid white curves show corresponding predictions including the pulse delay.
The additional oscillatory structure superimposing the data is due to the residual response of the incident SR. The bleached area $t\leq 50$~ns contains mostly data from this initial $\delta$-pulse excitation and is excluded from the data analysis.
(b) Sections through (a) at constant energies $\Delta_D$. Close to resonance $\Delta_D \approx 0$, the temporal response is clearly shifted compared to the off-resonant case. For example, the minimum at $t\approx 60$ ns is shifted to later times as indicated by the arrow. Solid lines are theoretical predictions for the pulse part $R_\textrm{SNXP}$ only, which is expected to deviate from the experimental data at initial times due to the omission of $R_\delta(t)$.
}
\end{figure}

{\it Experiment.} We performed the experiment at the nuclear resonance beam line (ID18)~\cite{id18} at the European Synchrotron Radiation Source (ESRF, Grenoble) using the setup sketched in Fig.~\ref{fig:implementation}. The x-ray cavity consists of a 
Pd(2\,nm)/C(20\,nm)/$^{57}$Fe(3\,nm)/C(21\,nm)/Pd(10 nm)/Si
layer system which is probed in grazing incidence such that the fundamental guided cavity mode is resonantly excited~\cite{Heeg2013b}. The high purity x-ray polarimeter is described in more detail in \cite{Marx2011}. A magnetic field is applied along the beam propagation direction, defining the quantization axis for the magnetic hyperfine splitting in the $^{57}$Fe layer. In this setting, vacuum-mediated couplings between the different hyperfine levels arise, which lead to steep linear dispersion as in EIT systems~\cite{Heeg2013,Heeg2013b} such that large time delays $\tau$ are expected. Note that in contrast to previous experiments focusing on the measurement of the absorption spectra~\cite{Roehlsberger2012,Heeg2013}, here, full transparency of the medium on resonance is not desirable, as it would correspond to zero intensity in reflection, prohibiting a detection of the propagated pulse. Therefore, the cavity system is chosen such that steep dispersion is obtained while maintaining sufficient intensity in reflection direction to enable the pulse detection.
The additional stainless steel foil (${}^{57}$Fe${}_{55}$Cr${}_{25}$Ni${}_{20}$) with $^{57}$Fe for the SNXP generation with thickness $10~\mu$m was mounted on a Doppler drive, such that pulses with different central frequencies $\omega_\textrm{SNXP} = \omega_0 + \Delta_D$ could be generated.

Due to the narrow nuclear linewidth, the SNXP consists, on average, of less than one photon. Triggering data acquisition on the detection of a photon at the detector thus essentially leads to post-selection of single photon SNXP. In the experiment, we registered the photon time of arrival together with the Doppler drive velocity for each signal photon separately. This enables us to analyze the intensity of the light registered by the detector as function of the pulse center frequency and time, as shown in Fig.~\ref{fig:counts}. Clearly, the time spectra of near-resonant pulses ($\Delta_D \approx 0$) are delayed compared to those of the off-resonant pulses, which can be seen, e.g., from the shift of a beating minimum at $t\approx 60$~ns to later times, see arrow in Fig.~\ref{fig:counts}(b). This figure also shows that the SNXP structure remains essentially undistorted. The agreement between experimental data and theoretical predictions is very good. From the theoretical analysis, we could also identify the additional oscillatory structures superimposing the simple temporal shift of the registered intensity by $\tau$ predicted in Eq.~(\ref{eq:E_2}) as arising from residuals of $R_\delta(t)$ in the data.

\begin{figure}[t]
 \centering
 \includegraphics[scale=0.66]{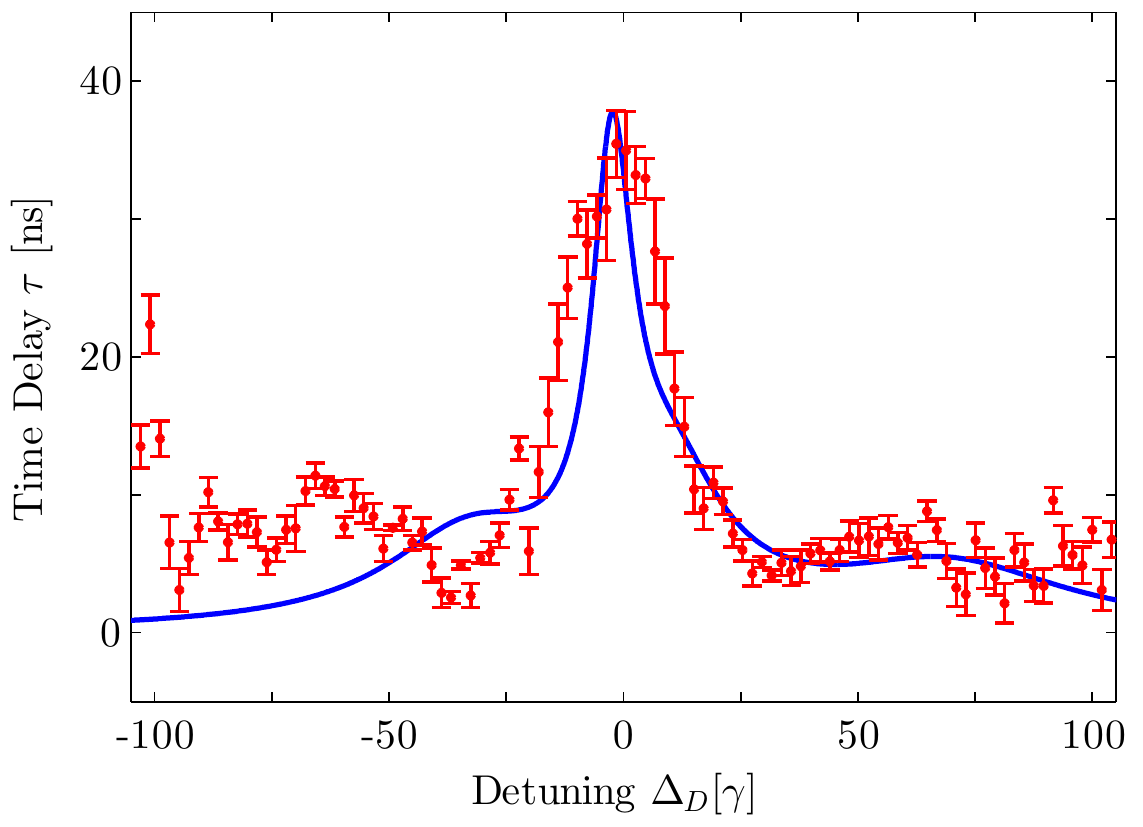}
 \caption{\label{fig:delay}(Color online) Time delay for the SNXP as function of the detuning $\Delta_D$ between SNXP and the nuclear resonance. Red dots show the delay extracted from the experimental data. The blue solid curve shows the corresponding theoretical prediction. Error bars are described in the \supplement.}
\end{figure} 

From the experimental data shown in Fig.~\ref{fig:counts}, we extracted the time delay $\tau$ of the x-ray pulses by fitting the analytical response function with variable $\tau$ to the data (Details on the employed fit method are provided in the \supplement). The result is shown in Fig.~\ref{fig:delay}.
As expected, around the cavity resonance where the nuclear susceptibility exhibits a steep positive linear dispersion, substantial pulse delays up to $35$~ns are observed. Away from the nuclear resonance, the delay reduces until it becomes zero off-resonance.

We have thus demonstrated group velocity control for spectrally narrow hard x-ray pulses, yielding controllable pulse delays of up to 35~ns via sub-luminal light propagation. The sub-luminal light propagation was realized by tailoring a suitable steep linear dispersion in M\"ossbauer nuclei embedded in an x-ray cavity. Our theoretical analysis agrees well with the experimental results. From numerical finite-difference time-domain simulations of the x-ray pulse dynamics, we determined an upper bound for the propagation length of the pulse inside the cavity of about $1$ mm, which translates into an upper bound for the reduced group velocity of the SNXP of $v_\textrm{gr} < 10^{-4}c$.
The group velocity control was enabled by a powerful method to generate SNXP, which requires neither mechanical choppers nor a high-resolution monochromatization of the incident SR light, and thus provides a route towards nuclear quantum optics experiments also beyond dedicated nuclear resonance beam lines. 

These results pave the way for a number of promising future directions. Our theory predicts that a suitable modification of sample magnetization and x-ray polarization offers means to tune the SNXP group velocity and thus the time delay, even to superluminal light propagation~\cite{Chu1982,Longhi2002}. Also a storage of x-ray photons could be envisioned. This way, slow light, EIT and related phenomena in the future could enable the coherence-based enhancement of non-linear interaction between x-rays and nuclei~\cite{Fleischhauer2005}. Next to this primary goal, our setup in turn could also be used to measure the phase of the response of an unknown sample, since the observed time delay is directly related to the phase of the optical response of the cavity-nuclei system.

K.P.H. acknowledges funding by the German National Academic Foundation and G.G.P. funding by Deutsche Forschungsgemeinschaft in the framework of CRC TR18.

\bibstyle{apsrev4-1}
%

\newcommand{\figcounts}{3 }

\newcommand{\figdelay}{4 }

\clearpage

\section*{Supplemental material}

\section{Pulse generation}
\subsection{Transmission function of the \textsuperscript{57}Fe foil}
The impact of the $^{57}$Fe foil (M\"ossbauer drive) on the transmitted x-rays can be described with the transmission function~\cite{Kagan1979,Shakhmuratov2009}
\begin{align}
 T_\textrm{foil}(\Delta) &= \exp{\left(-\frac{ i L \gamma/4}{\Delta - \Delta_D + \tfrac{i}{2}\gamma} \right)} \;,
\end{align}
where $\Delta = \omega -\omega_0$ is the detuning from the nuclear resonance, $\Delta_D$ accounts for an additional detuning due to the Doppler shift of the moving drive, $L = \sigma_0 f_\textrm{LM} n d$ denotes an effective thickness consisting of cross section $\sigma_0$, Lamb-M\"ossbauer factor $f_\textrm{LM}$, number density of resonant nuclei $n$ and foil thickness $d$. In the time domain it reads
\begin{align}
 T_\textrm{foil}(t) &= \sqrt{2\pi}\delta(t) - \theta(t) e^{-\tfrac{\gamma}{2}t - i\Delta_D t} \sqrt{\frac{\pi L \gamma}{2 t}} J_1\left(\sqrt{L \gamma t}\right) \;.
\end{align}
Here $\delta(t)$ denotes the Dirac delta function, $\theta(t)$ the Heaviside step function and $J_1$ the Bessel function of first order.
\subsection{Transmission function of the \textsuperscript{57}Fe foil including a chopper}
Applying a time domain chopper to the signal transmitted from the $^{57}$Fe foil leads to the modified time evolution
\begin{align}
 &T_\textrm{chopper}(t) = \theta(t - \tau_\textrm{chop}) T_\textrm{foil}(t) \nonumber \\
 &\qquad = \theta(t- \tau_\textrm{chop}) e^{-\tfrac{\gamma}{2}t - i\Delta_D t} \sqrt{\frac{\pi L \gamma}{2 t}} J_1\left(\sqrt{L \gamma t}\right) \;.
\end{align}
Transforming this expression back into the frequency domain yields
\begin{align}
 T_\textrm{chopper}(\Delta) &= \sum_{n=1}^\infty \frac{1}{n!} \left( \tfrac{-i L \gamma/4}{\Delta - \Delta_D + \tfrac{i}{2}\gamma} \right)^n 
 \nonumber \\
& \times \frac{\Gamma\left( n, [\tfrac{\gamma}{2} - i (\Delta - \Delta_D)] \tau_\textrm{chop} \right)}{(n-1)!}\;,
\end{align}
where $\Gamma(n, z) = (n-1)! e^{-z} \sum_{m=0}^{n-1} \tfrac{z^m}{m!}$ is the incomplete Gamma function. In the limit of small $\tau_\textrm{chop}$, the expression reduces to
\begin{align}
 T_\textrm{chopper}(\Delta) \approx T_\textrm{foil}(\Delta) - 1 \;,
\end{align}
which already represents a SNXP, since $T_\textrm{chopper}(\Delta)$ vanishes at larger detunings. Numerically, we find that a pulse is also obtained for general $\tau_\textrm{chop}$~\cite{Toellner2011}. Qualitatively, its spectral shape is rather unaffected by the temporal chopping. However, its amplitude strongly depends on the exact choice of $\tau_\textrm{chop}$.

\section{Group velocity control}
\subsection{Complex reflection coefficient of the cavity}
The reflectance of the cavity can be calculated analytically with the formalism developed in Ref.~\cite{Heeg2013b}. In our setup the polarimeter is operated in crossed setting, i.e.~the axes of the incident beam and of the analyzer are perpendicular. Furthermore, the magnetic field is oriented along the propagation direction. This configuration is known as Faraday geometry. The complex reflection coefficient reads~\cite{Heeg2013}
\begin{align}
 R_\textrm{Cavity} &= R_F(F_+) - R_F(F_-) \;, \label{rcav}
\end{align}
with the abbreviations
\begin{align}
 F_+ =&\frac{1/3}{\Delta + (-\tfrac{1}{2}\delta_g + \tfrac{1}{2}\delta_e) + i\tfrac{\gamma}{2}}
  \nonumber \\
 &+
       \frac{1}  {\Delta + (+\tfrac{1}{2}\delta_g + \tfrac{3}{2}\delta_e) + i\tfrac{\gamma}{2}} \,,\\
 F_- =&\frac{1  }{\Delta + (-\tfrac{1}{2}\delta_g - \tfrac{3}{2}\delta_e) + i\tfrac{\gamma}{2}} 
 \nonumber \\
 &+
       \frac{1/3}{\Delta + (+\tfrac{1}{2}\delta_g - \tfrac{1}{2}\delta_e) + i\tfrac{\gamma}{2}} \,,\\
 R_F(F) =& \frac{\kappa_R |g|^2 N}{(\kappa + i\Delta_C)^2} \left(\frac{2}{F} + \frac{i |g|^2 N}{\kappa + i\Delta_C} \right)^{-1} \;.
\end{align}
Here, $F_\pm$ are the nuclear scattering amplitudes, $\delta_g$ [$\delta_e$] is the energy difference between adjacent ground [excited] states with the values $22.4\gamma$ [$39.7\gamma$]~\cite{Hannon1999}, and $\gamma = 4.7$~neV is the width of the $^{57}$Fe transition. The coupling strength between cavity and the nuclei is denoted by $g$, the number of nuclei is $N$. $\kappa$, $\kappa_R$ and $\Delta_C$ are the cavity width, the coupling strength between the external x-ray and the cavity field, and the detuning from the cavity resonance.

\subsection{Time-resolved intensity at the detector}
In the frequency domain the signal arriving at the detector can be described via the product of the cavity reflection coefficient and the transmission function of the M\"ossbauer drive
\begin{align}
 I(\Delta, \Delta_D) \propto |R_\textrm{Cavity}(\Delta) \cdot T_\textrm{foil}(\Delta,\Delta_D)|^2\;.
\end{align}
Transforming the amplitude into the time domain yields the time-resolved signal at the detector, which we recorded in our experiment:
\begin{align}
 R(t, \Delta_D) &= \frac{1}{\sqrt{2\pi}} \int_{-\infty}^{\infty} R_\textrm{Cavity}(\Delta) \, T_\textrm{foil}(\Delta,\Delta_D) \, e^{-i \Delta t} d\Delta\,, \\
 I(t, \Delta_D) &\propto |R(t, \Delta_D)|^2 \;. \label{eq:intensity}
\end{align}
To evaluate the Fourier integral, we expand the transmission function
\begin{align}
 R(t, \Delta_D) =& \sum_{n=0}^\infty \frac{(-i L \gamma /4 )^n}{n!} (2\pi)^{-\tfrac 12}  \int R_\textrm{Cavity}(\Delta) 
 \nonumber \\
 &\times  (\Delta - \Delta_D+i\gamma/2)^{-n} \, \, e^{-i \Delta t} d\Delta  \nonumber \\
 =& R_\delta(t) + R_\textrm{SNXP}(t, \Delta_D)\,. \label{eq:R1_R2}
\end{align}
In the final step, we have split the sum into two parts, where $R_\delta$ covers the addend $n=0$ and $R_\textrm{SNXP}$ the rest. These two contributions correspond to the temporal responses of photons which did not interact with the M\"ossbauer drive foil [$R_\delta(t)$], and the desired signal of those photons which did interact [$R_\textrm{SNXP}(t, \Delta_D)$].

First, we calculate $R_\textrm{SNXP}(t, \Delta_D)$. Since the integral in each summand of $R_\textrm{SNXP}$ contributes mainly in the small range around $\Delta \approx \Delta_C$ we expand the cavity reflection coefficient. By assuming that its amplitude is constant and only the phase changes in this range, the approximation reads
\begin{align}
 R_\textrm{Cavity}(\Delta)& \approx R_\textrm{Cavity}(\Delta_D) \; e^{i (\Delta-\Delta_D) \, \frac{\partial \arg[R_\textrm{Cavity}]}{\partial \Delta} \Big|_{\Delta_D} } \nonumber \\[2ex]
 &= R_\textrm{Cavity}(\Delta_D) \; e^{i (\Delta-\Delta_D) \tau }\;, \label{eq:supp_R_cav_approx}
\end{align}
where we have defined the delay
\begin{align}
\tau = \frac{\partial \arg[R_\textrm{Cavity}]}{\partial \Delta} \Big|_{\Delta_D}  \,.
\end{align}
Inserting this expression into $R_\textrm{SNXP}$, we obtain
\begin{align}
 R_\textrm{SNXP}&(t, \Delta_D) 
\nonumber \\
\approx& \sum_{n=1}^\infty \frac{(-i L \gamma /4 )^n}{n!} (2\pi)^{-\tfrac 12} 
  \int R_\textrm{Cavity}(\Delta_D)  \nonumber \\
&  \times    (\Delta - \Delta_D+i\gamma/2)^{-n} \, e^{i (\Delta-\Delta_D) \tau } \, e^{-i \Delta t} d\Delta 
 \nonumber \\
 =&\sum_{n=1}^\infty \frac{(-i L \gamma /4 )^n}{n!} R_\textrm{Cavity}(\Delta_D) e^{-i\Delta_D t} (2\pi)^{-\tfrac 12} \nonumber \\
 &\times \int (\Delta +i\gamma/2)^{-n} \,  e^{- i \Delta (t - \tau) }  d\Delta \nonumber \\
 =& -  \sqrt{2\pi} e^{-\tfrac{\gamma}{2} (t-\tau)} \Theta(t-\tau)  R_\textrm{Cavity}(\Delta_D) \nonumber \\
 & \times  e^{-i\Delta_D t} \, \sqrt{\frac{L \gamma}{4 (t-\tau)}} J_1 \left(\sqrt{L \gamma (t-\tau)}\right)\,. \label{eq:R2}
\end{align}
We thus find that the signal is essentially only delayed by the time $\tau$ compared to the case without the cavity in the optical path ($R_\textrm{Cavity} = 1$, $\tau = 0$). Therefore, we see that the SNXP can indeed be delayed using the cavity without distortions.

Next we calculate $R_\delta(t)$, which corresponds to the cavity response to the $\delta$-pulse excitation. For this, we first consider
\begin{align}
 \tilde{R}(\Delta) = \left[ 2 \left( \frac{c_1}{\Delta + \delta_1 + i\tfrac{\gamma}{2}} + \frac{c_2}{\Delta + \delta_2 + i\tfrac{\gamma}{2}} \right)^{-1} + c_0  \right ]^{-1}
\end{align}
which has the same structure as each of the two addends in $R_\textrm{Cavity}$ in Eq.~(\ref{rcav}). Its Fourier transform is
\begin{align}
 \tilde{R}(t) &= \frac{1}{\sqrt{2\pi}} \int_{-\infty}^{\infty} \tilde{R}(\Delta) \, e^{-i \Delta t} d\Delta \nonumber \\
 &=\sqrt{\tfrac{\pi}{2}} \, e^{-\tfrac{\tilde{\Gamma}}{2} t} \, \Theta(t) \, \left[ 2\tfrac{\partial\tilde{\Omega}}{\partial c_0} \sin{\left( \tfrac{\tilde{\Omega} t}{2} \right) } \right . \nonumber \\
 & \left .\qquad  - i(c_1+c_2) \cos{\left(\tfrac{\tilde{\Omega} t}{2} \right)} \right] \label{eq:Rtildetrafo}
\end{align}
with the constants
\begin{align}
 \tilde{\Gamma} &= \gamma - \tfrac{i}{2} c_0 (c_1+c_2) - i (\delta_1 + \delta_2)\,, \\
 \tilde{\Omega} &= \left[ (\delta_1 - \delta_2)^2 + (\tfrac{c_0}{2})^2 (c_1+c_2)^2 \right . \nonumber \\
 &\left . \qquad + c_0 (\delta_1 - \delta_2) (c_1 - c_2) \right]^{\tfrac{1}{2}} \;.
\end{align}
Eq.~(\ref{eq:Rtildetrafo}) can easily be verified by transforming this expression back into the frequency domain. Using this result for both of the addends in $R_\textrm{Cavity}(\Delta)$, the Fourier transform $R_\delta(t)$ can be obtained in a straightforward way. Numerically, it can be seen that each term in $R_\delta(t)$ decays with a rate larger than $\gamma$, while $R_\textrm{SNXP}(t, \Delta_D)$ decays with only $\gamma$ [c.f.~Eq.~(\ref{eq:R2})]. Thus, the requirement of a fast decaying response of the cavity is met and for large $t$ only $R_\textrm{SNXP}(t, \Delta_D)$ determines the signal at the detector. Restricting the analysis to this time range therefore allows us to extract information on the SNXP only. For small $t$ both the responses of the SNXP and the $\delta$-pulse appear in the signal and their interference gives rise to the oscillating structures visible in Fig.~\figcounts of the main text.

\subsection{Relation to the susceptibility}
From slow light experiments in atomic media it is known that the group velocity and hence the time delay is related to the susceptibility of the medium~\cite{Harris1992,Fleischhauer2005}. In this part we will briefly derive this relation and compare it to our x-ray analysis.

Let us assume a spectrally narrow pulse $E(\omega)$, centered around the frequency $\omega_0$, which propagated through a medium of length $L$ and with refractive index $n(\omega) = \sqrt{1+\chi(\omega)} \approx 1 + \chi(\omega)/2$. It can be described by
\begin{align}
 E(L,t) = \frac{1}{\sqrt{2\pi}}\int E(\omega) e^{i(k L -\omega t)} d\omega \;,
\end{align}
with complex wave vector $k = n(\omega) \omega/c = k_R + i k_I$. In a typical atomic medium the susceptibility $\chi$ allows for the expansion
\begin{align}
 k_R(\omega) &\approx k_R(\omega_0) + \frac{\partial k_R}{\partial \omega}\Big|_{\omega_0}(\omega-\omega_0) \;,
 k_I(\omega) &\approx k_I(\omega_0) \;.
\end{align}
Then, the field becomes
\begin{align}
 E(L,t) &= \frac{1}{\sqrt{2\pi}} e^{-k_I (\omega_0) L} e^{-i (\omega_0 t - k_R(\omega_0) L)} \nonumber \\
 &\times \int E(\omega) e^{i(\omega-\omega_0) \left(\frac{\partial k_R}{\partial \omega}\big|_{\omega_0} L - t \right)}  d\omega\;, \label{eq:supp_E_slowlight}
\end{align}
where the first exponential accounts for absorption, the second for a global phase velocity and the integral covers the pulse envelope propagating with the group velocity $v_\textrm{Gr} = (\tfrac{\partial k_R}{\partial \omega}|_{\omega_0})^{-1}$.

Now let us turn to the x-ray reflection in our cavity setup. Here, the reflected pulse can be described by
\begin{align}
 E(L,t) = \frac{1}{\sqrt{2\pi}}\int E(\omega) R_\textrm{Cavity}(\omega) e^{i(k L -\omega t)} d\omega \;,
\end{align}
with the wavevector $k = \omega/c$. The vacuum dispersion relation can be used here since the full cavity response is already captured in $R_\textrm{Cavity}(\omega)$. With the approximation from Eq.~(\ref{eq:supp_R_cav_approx}), which states that the absolute value of $R_\textrm{Cavity}(\omega)$ is constant around $\omega_0$ and only the phase changes, we obtain
\begin{align}
 E(L,t) &= \frac{1}{\sqrt{2\pi}} R(\omega_0) e^{-i \omega_0 ( t - \tfrac{L}{c}) } \nonumber \\
 &\times \int  E(\omega) e^{i (\omega - \omega_0) \left(\frac{\partial \operatorname{arg}(R)}{\partial \omega}\big|_{\omega_0} + \frac{L}{c} -t \right)} d\omega \;.
\end{align}
Again, the first line describes absorption and global phase changes, while the integral covers the propagation of the pulse envelope. Comparing it to the envelope integral in Eq.~(\ref{eq:supp_E_slowlight}) and noting that $c \, \tfrac{\partial k_R}{\partial \omega} \big|_{\omega_0} = n_R(\omega_0) + \tfrac{\partial n_R}{\partial \omega} \big|_{\omega_0} \omega_0$, we can identify
\begin{align}
 \frac{\partial \operatorname{Re}(\chi)}{\partial \omega} \Big|_{\omega_0}
 \sim
 \frac{2 c}{\omega_0 L} \: \frac{\partial \operatorname{arg}(R)}{\partial \omega}\Big|_{\omega_0}
\;. \label{eq:supp_chi_re}
\end{align}
Similar, from the comparison of the absorptive parts we find
\begin{align}
 \operatorname{Im}\left(\chi(\omega_0)\right)
 \sim
 - \frac{2 c}{\omega_0 L} \: \log\left(|R(\omega_0)|\right) \;. \label{eq:supp_chi_im}
\end{align}
From these relations we can directly see that the phase and the modulus of the complex reflection coefficient take the role of the real and imaginary part of the susceptibility, respectively. Hence, a direct mapping between the theories for light propagation in atomic gases and for nuclear reflection is obtained.

Finally, we note that relations (\ref{eq:supp_chi_re}) and (\ref{eq:supp_chi_im}) are directly found by comparing
\begin{align}
 \exp{\left(i \,\frac{\omega_0 L}{2 c}\, \chi\right)}
 \sim
 R \;.\\[2ex]
 \nonumber
\end{align}

\section{Details on the fit method}
To determine the free parameters of our theory from the experimental data shown in Fig.~\figcounts in the main text, we minimized the deviation from the recorded data and the theoretical values calculated numerically with Eq.~(\ref{eq:intensity}). To account for the steep gradient along the time axis, the intensities were normalized along the detuning axis for each given time step. The best agreement was found for the cavity parameters $\kappa=45\gamma$, $|g|^2N = 3285\gamma^2$, $\Delta_C=-28.1\gamma$ and the effective thickness $L=126.3$ corresponding to a foil with thickness $10 \mu$m enriched to $85.4\%$ in $^{57}$Fe in the M\"ossbauer drive.

In a second step the time delay was determined by fitting the analytic expression for the time spectrum using Eqs.~(\ref{eq:R1_R2}) and (\ref{eq:R2}) to the data for each Doppler detuning $\Delta_D$. In this analysis, the cavity parameters determined above and a global scaling factor were kept constant, such that the only free parameter is the time delay $\tau$ entering Eq.~(\ref{eq:R2}). In order to extract the delay of the SNXP only, we suppress the contribution of the incident $\delta$-like pulse by restricting the fit range to times $t \ge 50$ ns. 
We found that due to the oscillatory structure of the data, the fit result can be affected by the starting value chosen for $\tau$. To extract unbiased values for $\tau$ from the data, we employed the following method: First, the best fit $\tau_0$ over 50 equidistantly distributed initial values in the range from $\tau = -0.1/\gamma$ to $0.4/\gamma$ was determined. Second, we performed another 50 fits with initial values in the range $\pm 0.25/\gamma$ around the previously determined $\tau_0$. Third, from these fits only the ones with $-0.1/\gamma \le \tau \le 0.4/\gamma$ were kept, since values outside this range clearly indicate an artifact caused by the oscillatory structure. Finally, we weight each $\tau$ with the inverse of its fit's variance to take into account the fit quality. From this final set of time delays $\tau$ the mean value and its standard error were determined.

Small distortions in the experiment, such as imperfect magnetization of the $^{57}$Fe layer, can result in a difference of the actual and the theoretically predicted spectrum. Far off-resonance, where $R_\textrm{Cavity}(\Delta_D) \ll 1$, this can lead to a large relative error. Since the fit function $R_\textrm{SNXP}(t, \Delta_D)$ directly depends on the spectrum $R_\textrm{Cavity}(\Delta_D)$, its amplitude is affected by the same error. Since the global scaling factor was kept constant, the time delay $\tau$ obtained from the fit might be distorted. This explains the discrepancies to the theoretical predictions in Fig.~\figdelay in the main text for large detuning.

\end{document}